# Spin transport in non-degenerate Si with a spin MOSFET structure at room temperature

Tomoyuki Sasaki [1,#], Yuichiro Ando [2,3,#], Makoto Kameno [2,#], Takayuki Tahara [3], Hayato Koike [1], Tohru Oikawa [1], Toshio Suzuki [4] and Masashi Shiraishi [2,3,$]

1. Advanced Technology Development Center, TDK Corporation, Japan.
2. Graduate School of Engineering Science, Osaka University, Japan.
3. Department of Electronic Science and Engineering, Kyoto University, Japan.
4. AIT, Akita Prefectural Industrial Center, Akita, Japan.

\# These three authors contributed equally to this work.

$ Corresponding author (mshiraishi@kuee.kyoto-u.ac.jp)

**Abstract**

**Spin transport in non-degenerate semiconductors is expected to pave a way to the creation of spin transistors, spin logic devices and reconfigurable logic circuits, because room temperature (RT) spin transport in Si has already been achieved. However, RT spin transport has been limited to degenerate Si, which makes it difficult to produce spin-based signals because a gate electric field cannot be used to manipulate such signals. Here, we report the experimental demonstration of spin transport in non-degenerate Si with a spin metal-oxide-semiconductor field-effect transistor (MOSFET) structure. We successfully observed the modulation of the Hanle-type spin precession signals, which is a characteristic spin dynamics in non-degenerate semiconductor. We obtained long spin transport of more than 20 μm and spin rotation, greater than $4\pi$ at RT. We also observed gate-induced modulation of spin transport signals at RT. The modulation of spin diffusion length as a function of a gate voltage was successfully observed, which we attributed to the Elliott-Yafet spin relaxation mechanism. These achievements are expected to make avenues to create of practical Si-based spin MOSFETs.**

# I. INTRODUCTION

The ability to control the flow of electrons based on their spin angular momentum by using two ferromagnetic electrodes, separated by a spin channel, i.e., a spin valve, has led to significant breakthroughs in the study of spin transport phenomena and their applications, resulting in tremendous advances in the field of spintronics [1-5]. Various techniques, such as electrical, dynamical and thermal methods, for generating pure spin currents have opened up the possibility of spin-based information devices [6,7]. Metals and semiconductors are important materials in spintronics, both have been subjects of rigorous studies. Since the first demonstration of spin transport in a hot-electron transistor in 2007 [8], various approaches have been used to achieve spin accumulation in Si, including devices based on electrical non-local 4-terminal [9-14], non-local 3-terminal [15-18] and dynamical [19] methods. Consequently, successful demonstration of spin transport in degenerate n-type [11] and p-type Si [19] at room temperature (RT) has been experimentally achieved.

Silicon is an attractive material for spintronics for the following reasons: (1) it has lattice inversion symmetry, resulting in good spin coherence. This is very advantageous when creating spin transistors (so-called "Sugahara-Tanaka" type spin transistors [3]); (2) Si is ubiquitous and non-toxic, which makes it an environment-friendly material that is important for the realization of a greener society; (3) Si-based electronics is already well established, and the enormous range of existing technologies and infrastructures can be fully utilized for mass-producing Si-based spin devices. Hence, Si is now regarded as one of the most promising materials for pushing technology beyond the complementary metal-oxide-semiconductor (CMOS). To accelerate the development of Si spintronics, it is essential to achieve long-range spin transport in non-degenerate Si at RT. Another important goal is the manipulation and modulation of spin signals using an external magnetic field and a gate voltage. This is because Si spin transistors, which would allow spin logic circuits to be fabricated [2,3], are based on the concept of conventional CMOS devices, in which source, drain and gate electrodes are used with a non-degenerate charge carrier channel. By using a hot-electron transistor structure, spin transport has been demonstrated at 260 K in non-generate n-type Si (resistivity 1-10 $\Omega$ cm) [20], and a spin transport length of 2 mm was achieved in intrinsic Si at 85 K [21]. However, this device structure is not considered appropriate for practical applications, because the output signal is on the order of several picoamperes, and the device itself is not easily integrated. It would therefore be difficult to fabricate spin-based logic circuits using this approach.

The purpose of this study was to investigate the possibility of controlling spin transport in non-degenerate n-type Si at RT by using spin metal-oxide-semiconductor field-effect-transistor (MOSFET) structure. Modulation of spin signals, in addition to spin diffusion length, is achieved by application of a gate electric field. A spin rotation of greater than $4\pi$ is realized by using an external magnetic field, and the results are well reproduced by a spin drift diffusion model. A spin transport length of more than 20 $\mu$m is demonstrated. The achievement of long-range spin transport at RT and the ability to easy manipulate spin signals using external fields in non-degenerate Si are expected to pave the way for the development of practical Si-based spin MOSFETs.

**II. EXPERIMENTAL**

The spin transport experiments in the present study were carried out using an electrical method. We have previously developed a variety of experimental techniques for observing spin transport signals, including non-local 4-terminal [10,11], modified non-local 3-terminal [22,23] and local 2-terminal [22] methods. Successful transport of a pure spin current and a spin-polarized current was achieved in degenerate n-type Si, and the results were supported by the observation of Hanle spin precession and magnetoresistance (MR). The same experimental platforms were also used in the present study. Spin-dependent transport and gate-induced modulation of the spin signals in the non-degenerate Si were measured using a 4-probe system (Janis ST-500) and a Physical Property Measurement System (PPMS, Quantum Design) at a temperature of 300 K. An external magnetic field was applied parallel and perpendicular to the Si channel plane in order to induce a MR effect and Hanle spin precession, respectively. For all measurements, the direction of spin flow was from contact 2 (pinned layer) to contact 3 (free layer) as shown in Fig. 1. The Si spin device was fabricated on a silicon-on-insulator substrate with the structure Si(100 nm)/SiO$_2$(200 nm)/bulk Si (see Fig. 1). The upper Si layer was phosphorous (P) doped by ion implantation. Using a 4-terminal method, the resistivity of the Si channel was determined to be 160 Ω μm, indicating that the dopant concentration was about $2 \times 10^{18}$ cm$^{-3}$ [24-31]. Before depositing the ferromagnetic electrodes, the Si in these regions was highly doped to a concentration of about $5 \times 10^{19}$ cm$^{-3}$, as determined by secondary ion mass spectroscopy [25]. Because of the insertion of the highly-doped region, the contact resistance decreases, resulting in the decrease of the whole resistance of the device. The decrease of the sample resistance allows efficient detection of spin signals. After the natural oxide layer on the Si channel was removed using a HF solution, Ti(3 nm)/Fe(13 nm)/MgO(0.8 nm) was grown on the etched surface by molecular beam epitaxy. Then, we etched out the Ti (3 nm)/Fe (3 nm) layers and Ta (3nm) was grown on the remaining Fe. In order to form FM contacts, the Si channel was etched to a depth of 25 nm by ion milling. The contacts had dimensions of 0.5×21 µm$^2$ and 2×21 µm$^2$, respectively. The Si channel surface and sidewalls at the FM contacts were buried by SiO$_2$. The nonmagnetic electrodes, with dimensions of 21×21 µm$^2$, were made from Al and were produced by ion milling. The gap between the FM electrodes was varied from 1.4 to 21.3 µm. The gate electric field was applied from the backside of the device.

## III. RESULTS AND DISCUSSION

Figure 2 shows the results of MR measurements carried out at 300 K. Local MR appeared to be present, with resistance hysteresis occurring between about ±200 and ±350 mT, as shown in Fig. 2(a). The sample resistance at 10 mA and the detected spin signal were 376 Ω and 170 µV, respectively, i.e., the MR ratio was 0.005%. In order to verify that the observed MR is related to the magnetization of the ferromagnetic (FM) electrodes, a modified 3-terminal method was used, as shown in Figs. 2(b) and 2(c). Clear MR was observed when the free FM layer was set as the base electrode ("MRF" configuration), but none was found when the pinned FM layer was set as the base electrode ("MRP" configuration). This is consistent with previous studies by our group [22,23], in which spin transport was achieved at RT in degenerate n-type Si. Here, we briefly summarize a principle of our modified non-local 3-terminal scheme (see also ref.22). The notable is that spin accumulation voltage is obtained from the ferromagnetic contact only under the spin extraction condition. The spin accumulation voltage obtained by the modified non-local 3-terminal measurement under the spin extraction condition is almost the same magnitude with that of local measurements in our previous study, indicating that the magnetoresistance in local scheme is caused by the spin accumulation voltage in one contact. This is the reason why the resistance hysteresis can be seen when the magnetization alignment of the detection FM electrode was reversed. Furthermore, as shown in Fig. 2(d), an apparent minor loop was observed, which was related to magnetization alignment of the FM electrodes. All of the above results provide evidence for successful spin transport in the non-degenerate n-type Si layer.

To obtain further confirmation of successful spin transport in the non-degenerate Si, Hanle-effect experiments were carried out. Hanle-type spin precession is not only an indicator of spin transport, but also of how easily the spins can be manipulated by an external field, in this case, a magnetic field. Figure 3 shows typical Hanle spin precession signals from a sample with a gap of 21.3 μm between the FM layers. The MRF configuration was used for the Hanle-effect experiment and the applied electric current was varied from 2 to 4 mA. A sample with a large gap was chosen in order to determine how many times the spins can rotate during transport. As shown in Fig. 3(a), symmetric oscillation of the spin signals as a function of the magnetic field was observed for both the parallel and anti-parallel magnetization configurations. Crossing of the parallel and anti-parallel signals occurred at about ±30 mT, indicating that the average spin rotation angle was π/2. It should be noted that spin rotation by up to 4π was realized, and such frequent rotation has previously only been found for intrinsic Si at low temperature [8]. In addition, a spin transport length of more than 20 μm was realized. In degenerated Si at RT, the spin transport length is limited to within 3 μm [10,11]. The spin drift diffusion equation is:

$$\frac{\partial S}{\partial t} = D\frac{\partial^2 S}{\partial x^2} - v\frac{\partial S}{\partial x} - \frac{S}{\tau}, \tag{1}$$

where $S(x,t)$ is spin density, $v$ is the spin drift velocity, $x$ is position, $t$ is time, and $\tau$ is the spin lifetime. This allows a quantitative estimation of the spin transport length. The analytic solution to Eq. (1) is [32]:

$$\frac{V(B)}{I} = \pm \frac{P^2 \sqrt{DT}}{2\sigma A} \exp(-\frac{L}{\lambda_N} + \frac{L}{2\lambda_N^2} v\tau)(1+\omega^2 T^2)^{-\frac{1}{4}} \times \exp[\frac{-L}{\lambda_N}(\sqrt{\frac{\sqrt{1+\omega^2 T^2}+1}{2}}-1)] \times \cos[\frac{\arctan(\omega T)}{2} + \frac{L}{\lambda_N}\sqrt{\frac{\sqrt{1+\omega^2 T^2}-1}{2}}], \quad (2)$$

where $P$ is the spin polarization, $A$ is the cross-sectional area of the channel, $L$ is the length of the gap between the two FM electrodes, $\omega = g\mu_B B/\hbar$ is the Larmor frequency, $g$ is the g-factor for the electrons ($g$ = 2 in this study), $\mu_B$ is the Bohr magneton, $\hbar$ is the Dirac constant, and the spin diffusion length is given by $\lambda_N = \sqrt{D\tau}$. Under spin drift, $T^{-1} = v^2/4D + 1/\tau$. The solid lines in Fig. 3(a) represent fits to the experimental data using Eq. (2). It can be seen that for both the parallel and anti-parallel configurations, the Hanle signals up to $4\pi$ are well reproduced using the spin drift diffusion model. Since the spin drift velocity was calculated to be 3720 m/s based on the resistivity, the spin lifetime and the spin diffusion length in the non-degenerate Si at RT were estimated to be 0.84 ns and 1.4 μm, respectively. Although the spin diffusion length is much shorter than the gap size in this sample, spin signals were successfully observed, which indicates the strong contribution of spin drift to spin transport.

Figure 3(b) shows the bias current dependence of the Hanle signals. It is clear that all of the peaks exhibit a monotonic shift to higher field with increasing bias current. No such shift was observed in the absence of a bias electric field in the spin channel in the non-local 4-terminal method. As shown in Fig. 3(b), the oscillations of the Hanle signals at 2, 3 and 4 mA are again well reproduced by the theory. The spin drift velocities at 3 and 4 mA are estimated to be 5570 and 7430 m/s, respectively. The good agreement between the experimental and theoretical results provides confirmation that the peak shift is due to the spin drift effect, and not to experimental errors.

Since successful spin transport was verified, an attempt was made to modulate the spin signal using a gate electric field. Figure 4 shows the gate voltage dependence of spin accumulation voltages that was detected in the MRF configuration as spin voltages, the channel resistivity, and MR ratio, respectively of the sample with the gap length of 1.7 μm ((a)-(c)) and 6.25 μm ((d)-(f)). The MR signal was almost constant for a negative gate voltage, but decreased monotonically with increasing positive gate voltage. Importantly, this behavior is qualitatively the same as that for the sample resistance as a function of the gate voltage, i.e., n-type FET characteristics are indicated. It is noteworthy that this is also consistent with the results of a theoretical analysis by Takahashi and Maekawa [33]. The MgO tunneling barrier is used in our spin device, resulting in the tunneling spin injection from the Fe to the non-degenerate Si. The spin voltage is, in theory, linearly proportional to the channel resistance in the tunneling spin injection, which is contrary to the Ohmic spin injection [33]. As shown in the figure, the spin voltage decrease as the channel resistance decreases by the gate voltage application, which is consistent with the theory. Thus, our experimental results indicate that the modulation of the spin voltages was successfully achieved by using the gate voltage application, which is due to both the non-degenerate Si spin channel and the existence of the tunneling barrier. Although the spin signal modulation is not large, probably due to the comparatively high doping level in the Si, this is the first demonstration of a Si-based spin MOSFET operating at RT.

The successful gate-voltage-induced modulation of the spin signals allows expecting the modulation of spin transport parameters, i.e., modulation of spin diffusion length and spin drift velocity as a function of the gate voltages. Figure 5 shows the gate voltage dependence of the Hanle signals of the sample with 6.25 μm gap, where the gate voltage was set to be -25, 0 and +50 V. As can be seen, the Hanle signals exhibit different behavior when the gate voltage was changed from -25 to +50 V. The theoretical fitting enables to estimate the spin diffusion length and the spin drift velocity, and it is clarified that they are modified as a function of the gate voltages as shown in Table I. The spin drift velocity decreases as the gate voltage increase because the electron accumulation, resulting in a decrease of resistivity, takes place under the positive gate voltage application. More importantly, the spin diffusion length also decreases under the positive gate voltage application. The electron accumulation induces an increase of a carrier density in the Si channel. Since the Elliott-Yafet type spin relaxation holds in Si [34], the increase of the carrier density enhances the spin relaxation, which attributes to the suppression of the spin diffusion length under the positive gate voltage application.

IV. SUMMARY

Spin transport at RT was successfully achieved in non-degenerate n-type Si using a spin MOSFET structure. A spin rotation of more than $4\pi$ was realized using an external magnetic field, and the results were well reproduced by a spin drift diffusion model. Long-range spin transport over a distance of more than 20 μm was also demonstrated. Modulation of the spin signal was shown to be possible using the gate voltage. The spin diffusion length was suppressed by the positive gate voltage application, which was attributed to the carrier accumulation, resulting in enhancement of the Elliot-Yafet type spin relaxation. These results are expected to open the door to the development of novel spin-based information devices using Si.

**ACKNOLEDGMENTS**

One of the authors (M.S.) thanks Prof. Y. Suzuki of Osaka University for his valuable suggestions and comments on the theoretical analyses. The part of this study performed by M.K. was supported by the Japan Society for the Promotion of Science (JSPS).

**Figure Captions**

**Figure 1.** Schematic structure of non-degenerate Si spin device. Electrodes 1 and 4 are nonmagnetic electrode made from Al, and electrodes 2 and 3 are ferromagnetic electrodes with MgO tunneling barriers. A highly-doped region was formed beneath the ferromagnetic electrodes, enabling efficient spin injection into the non-degenerate Si (see also ref. 26). The Si spin channel was formed on $SiO_2$. The gap length between the ferromagnetic electrodes, $L$, was varied from 1.4 to 21.3 μm, and the channel width was set to 21 μm.

**Figure 2.** Magnetoresistance (MR) effects in Si spin device. The schematics to the left of the experimental results indicate the circuit used for each measurement. Here, the gate voltage was not applied, and the external magnetic field was applied along y-axis. The red and blue solid lines in each panel represent the MR signals obtained under a forward and backward sweep of the external magnetic field, respectively. The bias electric current was 10 mA for each measurement, and the measurement temperature was 300 K. **(a)** MR signals observed using the electrical local 2-terminal method. Resistance hysteresis can be seen. **(b)** MR signals observed using the modified non-local 3-terminal method in the MRP configuration. No signal is seen in this configuration. **(c)** MR signals observed using the modified non-local 3-terminal method with the MRF configuration. Resistance hysteresis again appears. **(d)** Minor MR loop using the modified non-local 3-terminal method with the MRF configuration. The loop appears in the magnetic field region between two regions of resistance hysteresis. See the main text for details.

**Figure 3.** Hanle spin precession signals from non-degenerate Si without an application of a gate voltage. **(a)** Hanle signals for anti-parallel (blue open squares) and parallel (red open circles) magnetization configurations. Parabolic and linear background signals are subtracted from the raw data by taking the average of the signals and by linear fitting, respectively. The bias electric current was 2 mA and the measurement temperature was 300 K. The gap between the FM electrodes was 21.3 μm. A spin rotation of more than 4π could be observed. The blue and red solid lines are theoretical fits using Eq. (2), which reproduce the experimental results well. **(b)** Bias electric current dependence of the Hanle signals. The bias current was 2, 3 or 4 mA, corresponding to spin drift velocities of 3720, 5570 and 7430 m/s, respectively. Red, black and blue arrows show peak positions of π and 2π spin rotation at 2, 3 and 4 mA, respectively. Pale red, black and blue solid lines are theoretical fits using Eq. (2).

**Figure 4.** Si spin MOSFET operation at RT of the sample with 1.7 μm gap ((a)-(c)) and with 6.25 μm gap ((d)-(f)). (a) and (d) are channel resistivity (FET characteristics), (b) and (e) spin voltages, and (c) and (f) MR ratio. A similar dependence on the gate voltage is seen in both cases, which can be explained using the theoretical model of Takahashi and Maekawa [27].

**Figure 5.** Gate voltage dependence of the Hanle signals from the sample with 6.25 μm gap. The gate voltage was set to be -25, 0, +50 V. The open circles are experimental data and the solid lines are theoretical fitting lines.

**Table and Table caption**

Table I. Estimated spin drift velocity and spin diffusion length as a function of the gate voltages.

| $V_g$ (V) | -25 | 0 | 50 |
|---|---|---|---|
| Spin drift velocity (m/s) | 4480 | 2620 | 1570 |
| Spin diffusion length (μm) | 4.53 | 1.66 | 0.85 |

**Figures**

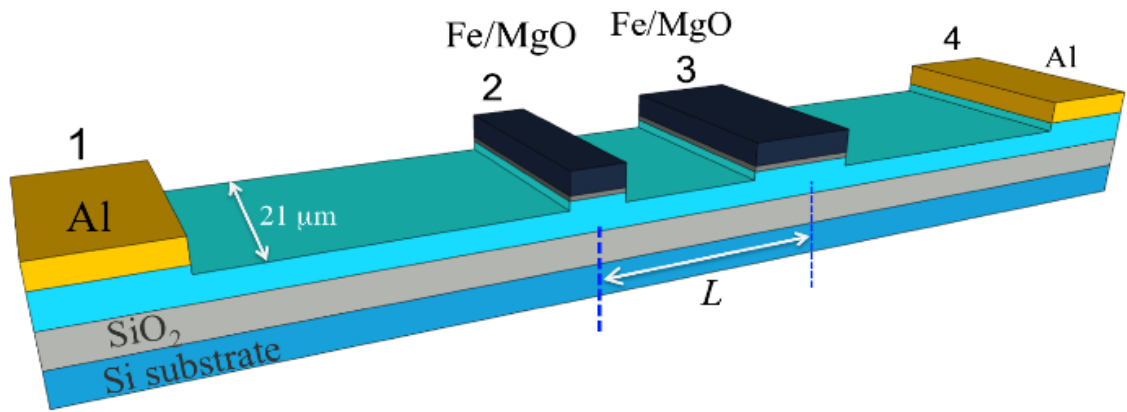

Figure 1 T. Sasaki et al.

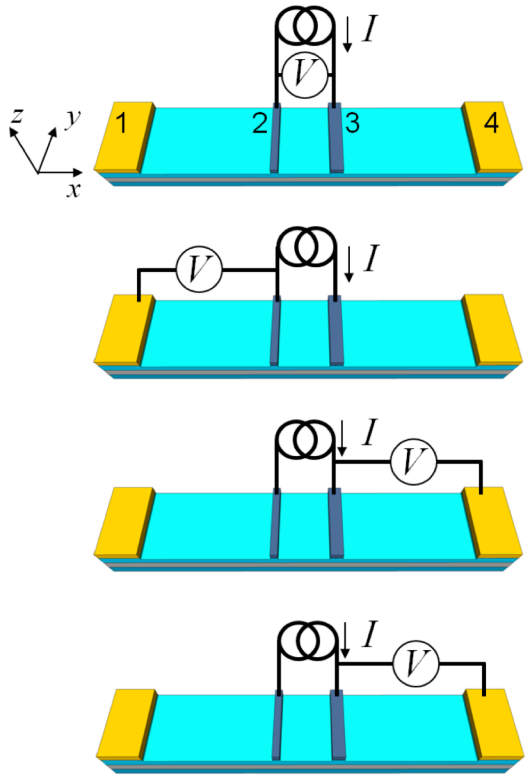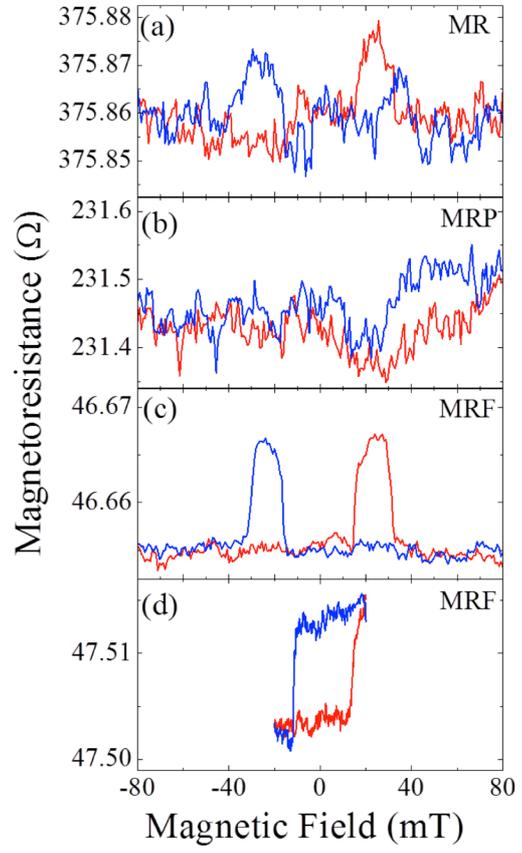

Fig. 2(a)-(d) T. Sasaki et al.

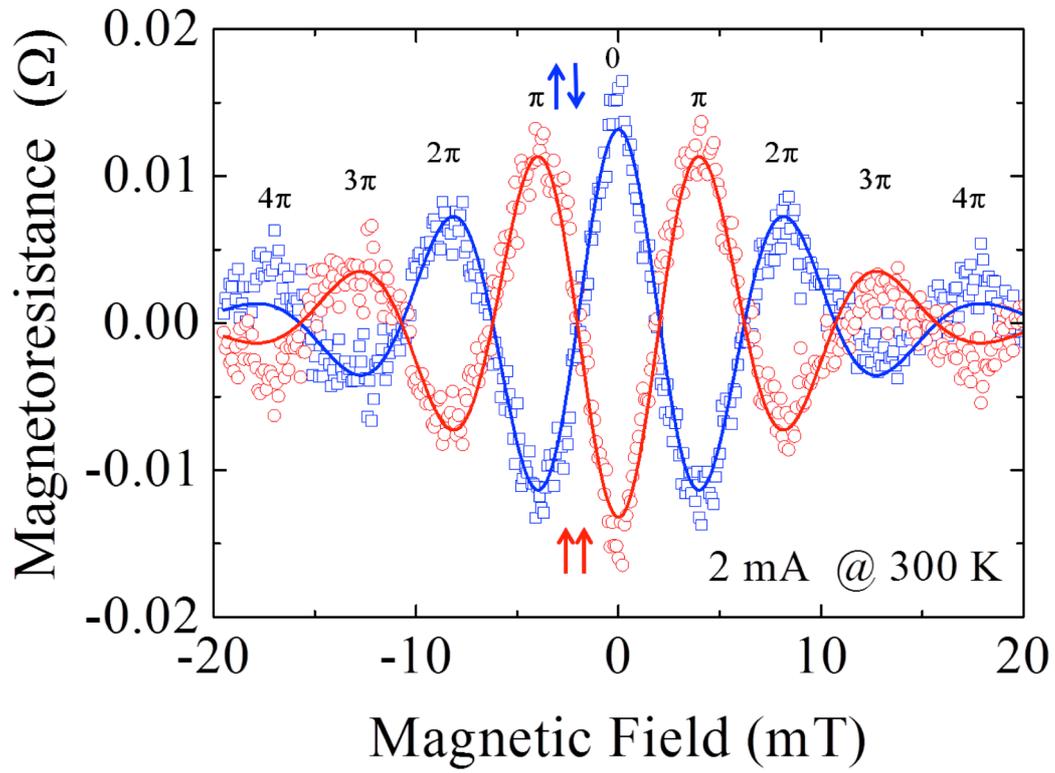

Fig. 3(a) T. Sasaki et al.

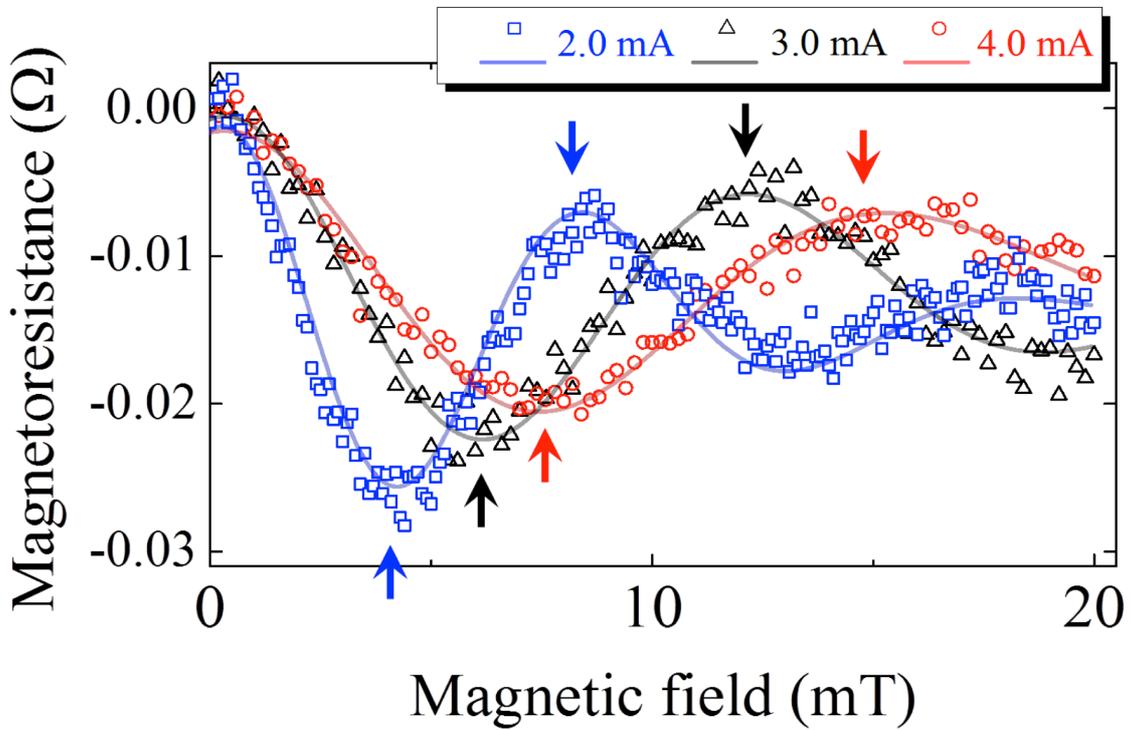

Fig. 3(b) T. Sasaki et al.

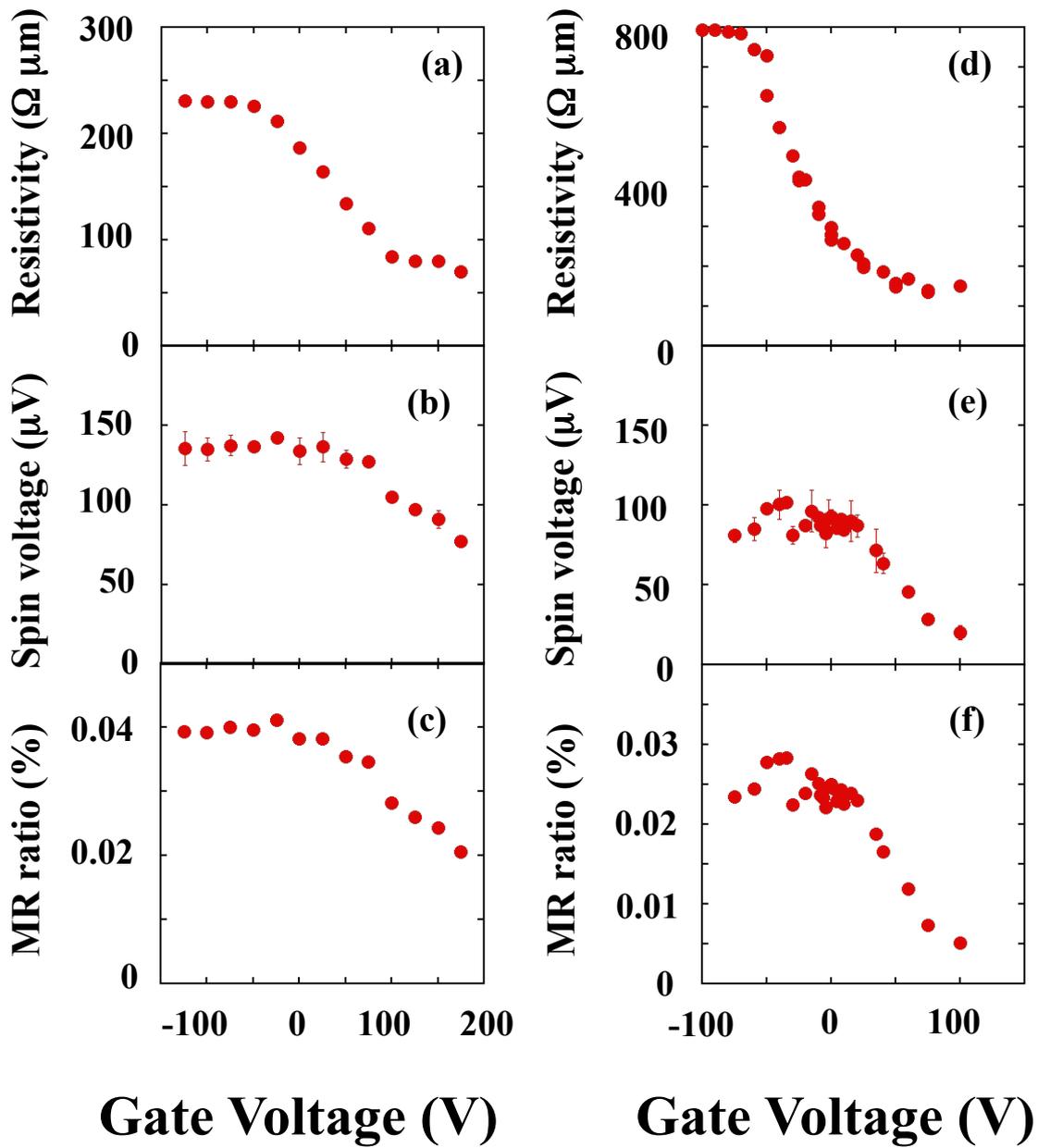

Fig. 4 T. Sasaki et al.

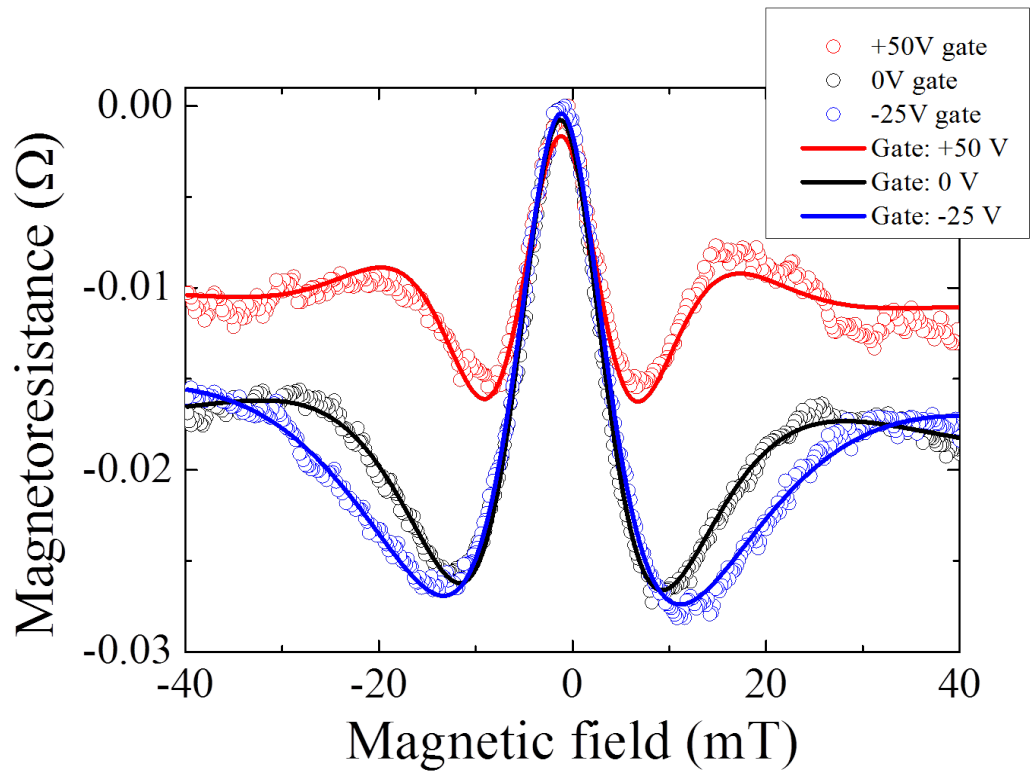

Fig. 5 T. Sasaki et al.